\documentclass[review]{elsarticle}

\usepackage{lineno,hyperref}
\modulolinenumbers[5]

\journal{High Energy Density Physics}

\usepackage{graphicx}
\usepackage{dcolumn}
\usepackage{bm}
\usepackage{amsmath}
\usepackage[caption=false]{subfig}

\begin{document}

\begin{frontmatter}

\title{Comparison of ablators for the polar direct drive exploding pusher platform}
\tnotetext[mytitlenote]{Manuscript prepared for the proceedings of IFSA2019.}


\author[myaddress2]{Heather D. Whitley\fnref{myfootnote1}}
\fntext[myfootnote1]{Presenting and corresponding author}
\ead{whitley3@llnl.gov}

\author[myaddress2]{G. Elijah Kemp}

\author[myaddress2]{Charles Yeamans}

\author[myaddress2]{Zachary Walters}

\author[myaddress2]{Brent E. Blue}

\author[myaddress3]{Warren Garbett}

\author[myaddress2]{Marilyn Schneider}

\author[myaddress1]{R. Stephen Craxton}

\author[myaddress1]{Emma M. Garcia}

\author[myaddress1]{Patrick W. McKenty}

\author[myaddress4]{Maria Gatu-Johnson}

\author[myaddress2]{Kyle Caspersen}

\author[myaddress2]{John I. Castor}

\author[myaddress2]{Markus D\"{a}ne}

\author[myaddress2]{C. Leland Ellison}

\author[myaddress2]{James Gaffney}

\author[myaddress2]{Frank R. Graziani}

\author[myaddress2]{John Klepeis}

\author[myaddress2]{Natalie Kostinski}

\author[myaddress2]{Andrea Kritcher}

\author[myaddress4]{Brandon Lahmann}

\author[myaddress2]{Amy E. Lazicki}

\author[myaddress2]{Hai P. Le}

\author[myaddress2]{Richard A. London}

\author[myaddress2]{Brian Maddox}

\author[myaddress2]{Michelle Marshall}

\author[myaddress2]{Madison E. Martin}

\author[myaddress5]{Burkhard Militzer}

\author[myaddress2]{Abbas Nikroo}

\author[myaddress2]{Joseph Nilsen}

\author[myaddress2]{Tadashi Ogitsu}

\author[myaddress2]{John Pask}

\author[myaddress2]{Jesse E. Pino}

\author[myaddress3]{Michael Rubery}

\author[myaddress2]{Ronnie Shepherd}

\author[myaddress2]{Philip A. Sterne}

\author[myaddress2]{Damian C. Swift}

\author[myaddress2]{Lin Yang}

\author[myaddress1]{Shuai Zhang}

\address[myaddress2]{Lawrence Livermore National Laboratory, Livermore, California 94550, USA}
\address[myaddress3]{AWE plc, Aldermaston, Reading RG7 4PR, United Kingdom}
\address[myaddress1]{Laboratory for Laser Energetics, University of Rochester, Rochester, New York 14623, USA}
\address[myaddress4]{Massachusetts Institute of Technology, Plasma Science and Fusion Center, Cambridge, Massachusetts 02139, USA}
\address[myaddress5]{University of California, Berkeley, California 94720, USA}

\begin{abstract}
We examine the performance of pure boron, boron carbide, high density carbon, and boron nitride ablators in the polar direct drive exploding pusher (PDXP) platform. 
The platform uses the polar direct drive configuration at the National Ignition Facility to drive 
high ion temperatures in a room temperature capsule and has potential applications for plasma physics studies and as a neutron 
source.  
The higher tensile strength of these materials compared to plastic enables a thinner ablator to support 
higher gas pressures, which could help optimize its performance for plasma physics experiments, while ablators containing boron enable the possibility of collecting additional data to constrain models of the platform.   
Applying recently developed and experimentally validated equation of state models for the boron materials, we examine the performance of these materials as ablators in 2D simulations, with particular focus on changes to the ablator and gas areal density, as well as the predicted symmetry of the inherently 2D implosion.
\end{abstract}


\end{frontmatter}

\section{Introduction}

The Polar Direct Drive Exploding Pusher (PDXP) platform was proposed and developed as a platform for studying 
electron-ion temperature equilibration and thermal conduction in the high energy density regime that is relevant to 
inertial confinement fusion at the National Ignition Facility (NIF)\cite{Ellison18, Miles_2012,Skupsky04}  It has since been applied in both nucleosynthesis experiments\cite{GatuJohnson18} and as a neutron source.\cite{Yeamans18,Yeamans20}  
  Our initial PDXP proposal for NIF called for a thin ablator, 
enabling full ablation of the capsule 
shell, which we believed would lead to better uniformity of the plasma during the proposed time-resolved 
spectroscopic measurements of the plasma temperature.  Early design studies indicated that the performance for heat flow 
measurements was optimized with a gas fill pressure of 8-10~atm based on 500~kJ of laser energy incident on a 3~mm outer diameter capsule.  
Because the proposed measurements of plasma temperature rely on using Ar as a spectroscopic dopant,  the platform required that the signal from the Ar spectral lines must be 
significantly higher than the emission from the background plasma, and the Ar mass in the target must be well known.  We had initially considered SiO$_2$ or Be ablators for these measurements due to the ability to fabricate thin capsules of 
either material.  The SiO$_2$ design 
was ruled out due 
to calculations that showed high background emission, and thus low Ar signal, during the proposed measurement, and Be 
was ruled out because the sputtering process used to make Be ablators generally results in significant Ar remaining in the shell.
For these reasons, and the lack of capabilities to build high density carbon (HDC) capsules 
of the desired size at the time, we based our point design on glow discharge 
polymer (GDP) ablators, which necessitated capsules of $\sim$20~$\mu$m thickness for the desired fill pressure.\cite{Nikroo04}  
The initial shots were thus fielded using 3~mm diameter GDP capsules with thicknesses 
of 18-20~$\mu$m and $\sim$8~atm gas fill.  The inflight implosion self-emission measurements and post-shot simulations from these initial  
shots (N160920-003, N160920-005, N160921-001, N170212-003, and N170212-004) indicated 
slight inflight asymmetry early in the implosion and a very asymmetric shell at 
bang time.\cite{Ellison18,GatuJohnson18,specialWhitley}

The laser pulse design in PDXP was motivated by the general concepts associated with the design of exploding pushers; the 
optimal design results from a rapid, impulsive ablation of the capsule, driving very high ion temperatures in the fill gas.\cite{Rosen79}  
The optimization of the pulse for the initial design of the heat flow platform was completed by examining a series of 1D radiation hydrodynamic 
simulations and choosing a pulse that optimized the window of time available for 
electron and ion temperature measurements.  This resulted in a 1.8~ns square pulse and computed ablator 
mass remaining of about 30\% at the end of the pulse.  All subsequent shots on this platform have similarly used pulse shapes where 
the majority of the laser energy is delivered during a square pulse, and 1D 
simulations of those shots indicate a similar amount of remaining mass, based on the total mass contained within the 
contour of electron density corresponding to the critical surface for laser absorption at 351~nm wavelength ($\sim9\times 10^{21}$~cm$^{-3}$), regardless of laser drive or capsule geometry.  Although these 
capsules are driven by relatively short laser pulses, 2D simulations show that the laser beams tend to continually imprint a specific 
pattern on the imploding shell, and this imprint appears to contribute to the observed capsule asymmetry at bang time 
based on comparison of the self-emission images from N170212-003 and N170212-004 to 2D simulations.\cite{GatuJohnson18}  

One possible route for mitigating the asymmetry, which would presumably allow for the generation of more uniform plasma conditions,  
would be to design capsules that have a thinner ablator with better coupling to the laser.  Such an ablator could potentially enable the 
use of a shorter pulse, and the higher thermal conductivity of a higher density material could help to mitigate the non-uniformity of the 
laser energy deposition.  Due to the linear relation between tensile strength and capsule burst pressure,\cite{Nikroo04} materials such as 
boron (B), high density carbon (HDC), boron carbide (B$_4$C), and boron nitride (BN), which have tensile strength 5-10 times higher than that of GDP,  
could presumably support the 8-10~atm fill pressures of the nominal PDXP point design at substantially reduced 
thickness relative to GDP.  While HDC is now a common capsule material, our interest in boron-containing materials is  motivated by the possibility of 
collecting data to help constrain simulation models of the PDXP platform.  In PDXP capsules with a DT gas fill, high yields can be 
achieved, and thus comparing gamma reaction history (GRH) measurements\cite{GRHref,GRH1} from implosions 
using an ablator containing natural boron to measurements using a GDP 
capsule could potentially provide constraining data for the gas areal density during burn due to the impact of 
knock-on deuterons on the 
$^{11}$B(d,n$\gamma_{15.1}$)$^{12}$C reaction on the GRH.\cite{Hayes15}  In addition, our best fitting simulations of previous shots invoke a diffusive mix model 
for ablator-fuel mix.\cite{Ellison18}  The $^{10}$B($\alpha$,p$\gamma$)$^{13}$C reaction, which produces $\gamma$ signals around 3.5~MeV, could provide data to 
help distinguish between diffusive mix and hydrodynamic instabilities, potentially validating the use of this diffusive mix model.\cite{Hayes17}
We note that our interest in the pure B and BN ablators is specifically motivated by the absence of carbon in these ablators, 
which eliminates potential cross talk from other reactions with C.\cite{specialHayes}  
These same reactions with C are useful for constraining shell areal density based on GRH data\cite{GRH2}, but would complicate the 
diagnostics we propose here for examining the gas density and distinguishing between diffusive mix and hydrodynamic instabilities.

Over the past several years, advances in additive 
manufacturing and target fabrication techniques have made the possibility of fielding shots with B$_4$C ablators 
more tangible.\cite{Chen17}  Novel techniques have also been applied to 
make targets for planar equation of state experiments on BN at the NIF.\cite{Dufrane16}  It therefore seems timely to examine these 
materials as potential ablators.  We are not currently aware of a fabrication technique for making a pure B capsule, though we include our results for B 
for future comparison purposes.
We present a brief summary of simulations examining the performance of B, B$_4$C, HDC, and BN ablators in 2D.  
Our 2D models are based on previously developed 
post-shot models for N160920-005, which fielded a GDP ablator and 8~atm D$_2$ gas fill at room temperature.  
Due to the inherent uncertainties in modeling capsule implosions, we seek to minimize controllable sources of error in this work.  
As a prelude to this study, we therefore applied a variety of theoretical methods to examine the equation of state (EOS) 
of pure boron,
B$_4$C, and BN\cite{Zhang2018b,Zhang2019BN,Zhang20} since these materials have not yet been used in capsule experiments at the NIF.  
New EOS models were developed for B and BN based on our earlier work, and we make use 
of a previously developed and recently tested model for B$_4$C\cite{Zhang20, Sterne_2016} in this baseline comparison study.  The EOS of HDC and GDP were also previously studied in 
detail.\cite{Sterne_2016,Zhang2018,Benedict2014}

\section{Model description and results}

\begin{figure}
\begin{center}
\includegraphics[width=0.75\textwidth]{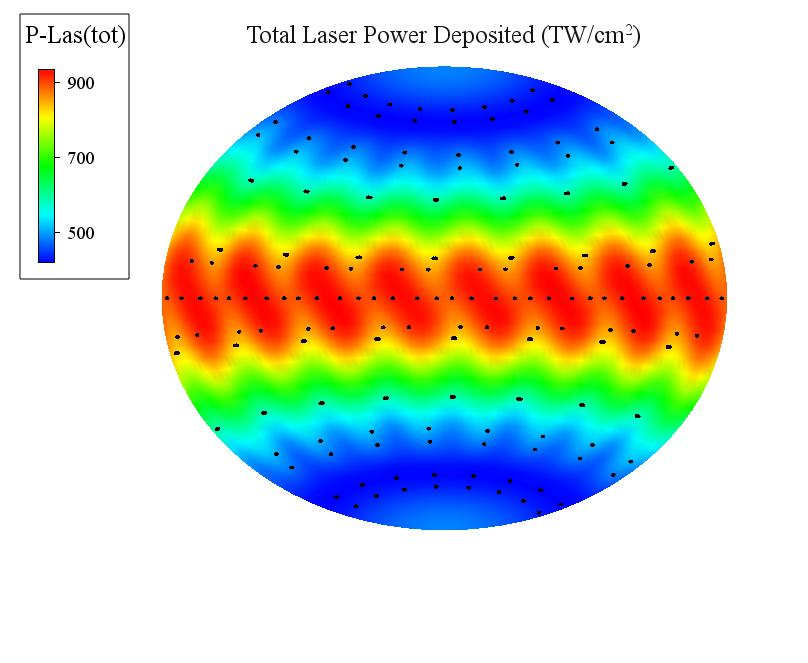}
\caption{Computed laser power on the capsule surface for N160920-005.  The black dots indicate the pointing on the capsule surface.}
\label{power}
\end{center}
\end{figure}

Our 2D direct drive simulations are carried out using the Ares radiation hydrodynamics simulation code.\cite{Darlington01,Morgan16}  For the purpose of this study, we use N160920-005 as the baseline for tuning the initial model and we 
use the laser pulse as delivered in this shot for all simulations reported here.  
In this shot, we fielded a 2.955~mm outer diameter GDP with a 19~$\mu$m thickness ablator, filled with 7.941~atm of $D_2$ gas with $5\times 10^{-4}$~atomic fraction of 
Ar as a spectroscopic dopant.  The capsule was driven with a 1.8~ns square pulse, delivering 479~kJ of total energy with slightly higher power in the 
outer beams to provide additional power near the equator of the capsule.\cite{Ellison18}  The calculated power profile on the capsule surface is 
shown in Figure~\ref{power}.
We use a laser ray trace method for depositing the energy in the capsule, which 
takes into account the 3D pointing geometry, but does not include the effects of cross-beam energy transfer or nonlocal electron thermal transport.  Both of these effects are known to be important for modeling 
laser-matter interactions in direct drive implosions,\cite{Krasheninnikova_2014, Murphy_2015, Radha16, Zylstra20} 
but we have nonetheless found that the salient features of our shots are modeled well using a more approximate treatment.  
Our models employ multi-group diffusion for the propagation of radiation, and we apply 
a flux limiter to the electron thermal conduction in the ablator during the laser pulse.
We tune the flux limiter and a multiplier on 
the total laser power to fit the observed x-ray bang time of the shot, as described in Ref.~\cite{Ellison18}.  In this study, we used a flux limiter of 0.0398, and we 
find a good fit to the neutron bang time by assuming an energy multiplier of 0.875.  
We have also applied the multicomponent Navier-Stokes (mcNS) model for species diffusion in simulations of this shot, 
and we find that using this model enables a good match to the measured burn-averaged ion temperature 
and the neutron yield, provided that a multiplier is applied to the diffusion coefficient.\cite{Ellison18}  
However, we have no reason to expect that the multiplier that we determined for the GDP capsules will also apply to the 
ablators considered in this study, and so we did not exercise the species diffusion model in this study. 

Table~\ref{capsparms} summarizes the ablator characteristics of the 2D simulations performed in this study.  
For HDC, we considered both a thin design and a thicker 
 design. For the thicker design, 
the ablator thickness was chosen to be 6.0~$\mu$m in order provide a mass 
match to the GDP ablator, whereas for the thinner designs, we first considered an HDC capsule where the 
total ablator mass is reduced to 0.4~mg, corresponding to a thickness of 4.45~$\mu$m.  
The thicknesses of the B and B$_4$C ablators were then chosen to match the total mass of the thinner HDC design (6.0~$\mu$m and 5.86~$\mu$m, 
respectively).  Similar to HDC, BN can exist either in a cubic (diamond) lattice or in a hexagonal (graphitic) lattice.  In this work, we consider BN in the 
hexagonal phase, with a density of 2.25~g/cc, so the mass of the 
BN ablator is just slightly lower than that of the other thin capsule designs.  The BN capsule was chosen to have a thickness that matches the thin HDC capsule.  

Table~\ref{capsparms} also lists the equation of state model used for each 
material in the table.  For BN, we applied both a model that was recently developed (L2152)\cite{Zhang2019BN} and an older model from the LEOS library 
that was developed by D.~A. Young and is based on a Thomas-Fermi model (L2150).  These two models were compared in our previous report on the BN 
equation of state.\cite{Zhang2019BN}  For each of these calculations, we assumed a D$_2$ fill pressure of 7.941~atm at room temperature, which was chosen to match N160920-005.  
We use the L1014 model for the EOS of D$_2$, consistent with our previous 1D simulation studies.\cite{Ellison18,Zhang2018b,Zhang20}

\begin{center}
\begin{table}
\begin{tabular}{ccccc}
\hline
Ablator & Thickness & Capsule Mass & Density & EOS Models\\
& ($\mu$m) &   (mg) & (g/cc) & \\
\hline
GDP & 19 &  0.54 & 1.046 & L5400\cite{Sterne_2016,Zhang2018} \\
HDC & 6 & 0.54 & 3.32 & L9061\cite{Benedict2014} \\
B & 6 & 0.40 & 2.46 & L52\cite{Zhang2018b} \\
B$_4$C & 5.86 & 0.40 & 2.52  & L2122\cite{Sterne_2016,Zhang20}\\
BN & 6 & 0.37 & 2.25 & L2152\cite{Zhang2019BN} and L2150\\
HDC & 4.45 & 0.40 & 3.32 & L9061\cite{Benedict2014} \\
\end{tabular}
\caption{Capsule parameters and EOS models used in this study.}
\label{capsparms}
\end{table}
\end{center}

Table~\ref{yieldtable} lists some of the computed results for each of the capsules.  
The total yield of the capsules predictably increases as a function of the total energy absorbed. Since the
laser-capsule coupling is higher for the higher density ablators, the HDC and BN ablators produce the highest neutron yields.  
We also  find that the two HDC capsules absorb the same amount of energy from the laser, 
but the thicker ablator produces higher yield, 
higher peak convergence ratio ($CR$) defined based on the ratio of the initial gas volume to the minimum gas volume, 
$CR=(V_i/V_{min})^{1/3}$, and  lower burn-averaged ion temperature.  The 
performance of the thicker HDC capsule appears to mimic that of the thick GDP capsule.   
Results for the BN capsule are reported only for L2152 because the results from L2150 were nearly identical.  
As expected, the thin capsule with low ablator density near peak compression is not sensitive to the choice of EOS model.  (For the thicker capsules, 
variations in the EOS can impact the computed performance, and we explore EOS variations in greater depth in Ref.~\cite{Zhang20}.)

\begin{center}
\begin{table}
\begin{tabular}{cccccc}
\hline
Ablator/thickness ($\mu$m) & Absorbed &Neutron & Convergence &  $T_{ion}$  \\
& Energy (kJ) & Yield & Ratio &(keV) \\
\hline
GDP/19 & 282 & $4.4\times 10^{13}$ & 9.8 & 7.2\\
HDC/6 & 324 & $9.1\times 10^{13}$ & 10 & 8.8 \\
B/6 & 310 & $6.2\times 10^{13}$ & 5 & 15\\
B$_4$C/5.86 & 313 & $6.5\times 10^{13}$ & 5 & 15\\
BN/6 & 319 & $6.7\times 10^{13}$ & 5 & 16\\
HDC/4.45 & 324 & $7.8\times 10^{13}$ & 5 & 15\\
\end{tabular}
\caption{Results from 2D Ares simulations.  The convergence ratio is computed based on the minimum gas volume.  We note that the 
measured neutron yield from N160920-005 was $2.11(\pm 0.1) \times 10^{13}$, so the clean yield computed in the 2D calculation of the GDP capsule is about a factor of 2 larger 
than the experiment.}
\label{yieldtable}
\end{table}
\end{center}

\begin{figure}
\begin{center}
\includegraphics[width=0.9\textwidth]{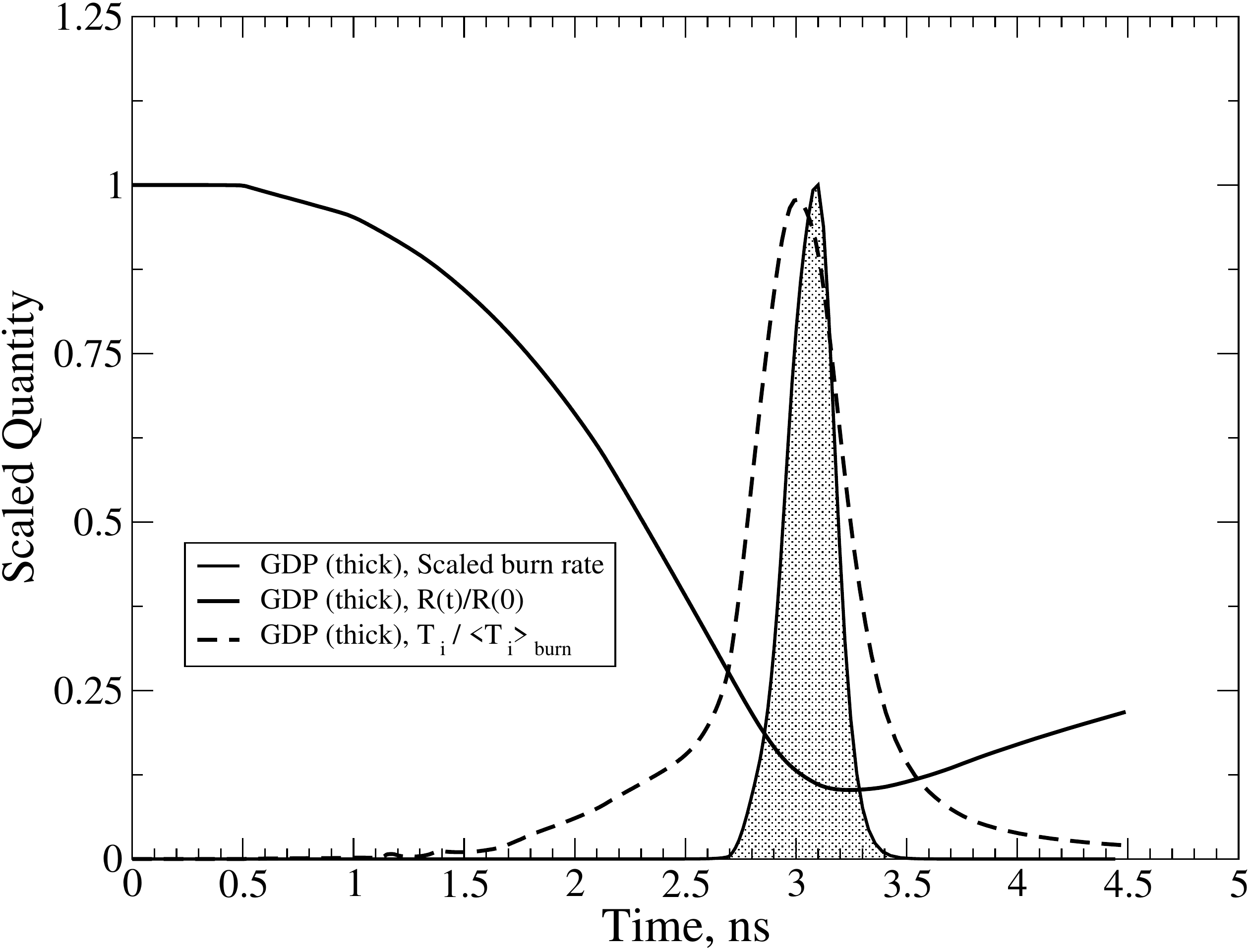}
\caption{Scaled burn rate (shaded curve), average radius (solid), and average ion temperature (dashed) as a function of time for the GDP design, as described in the text.  The burn in the thicker GDP design takes place primarily after the peak average temperature in the gas is reached, implying that 
compression of the gas is contributing to the overall yield.}
\label{GDPpdxp}
\end{center}
\end{figure}

\begin{figure}
\begin{center}
\includegraphics[width=0.9\textwidth]{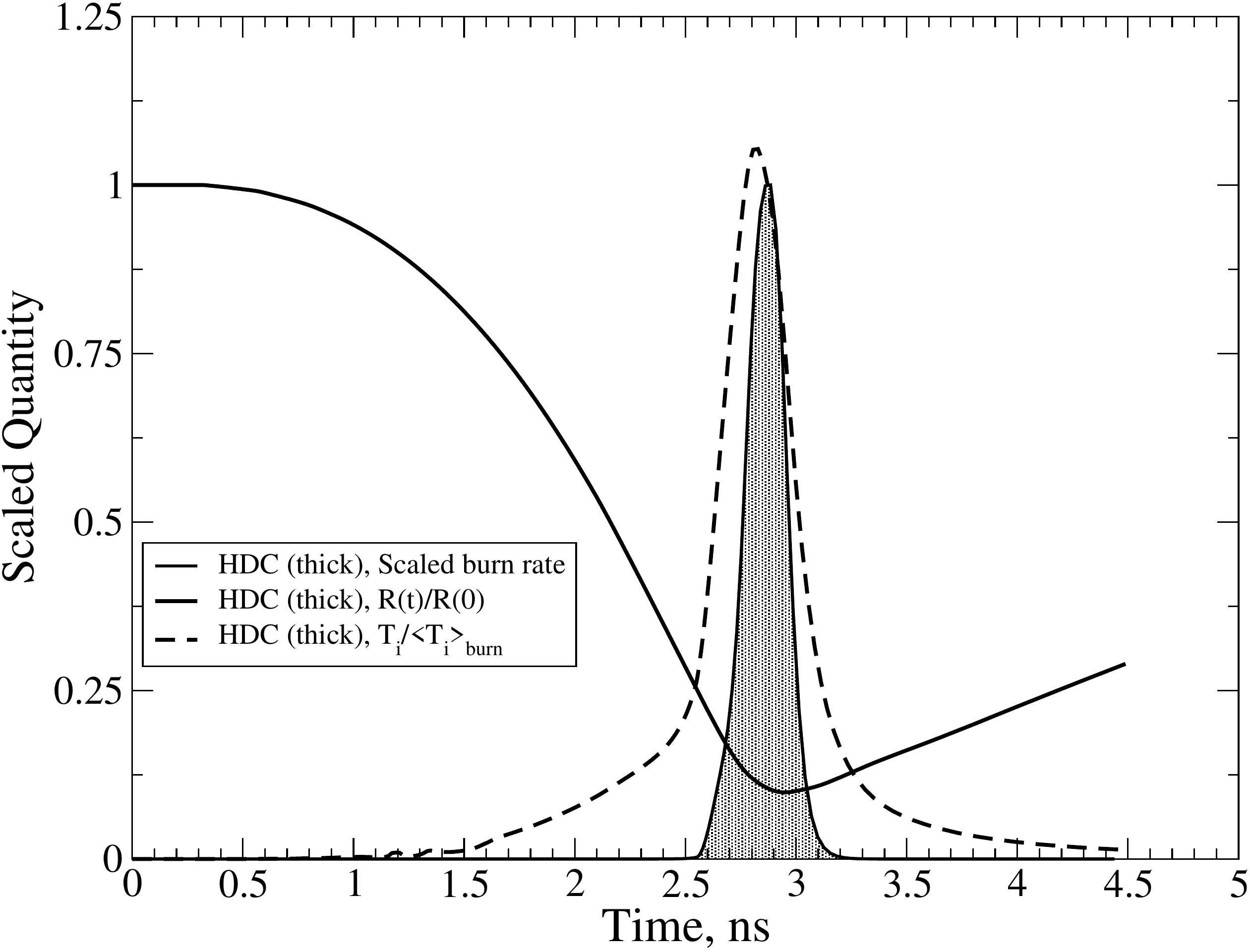}
\caption{Scaled burn rate (shaded curve), average radius (solid), and average ion temperature (dashed) as a function of time for the thicker HDC design, as described in the text.  Similar to the GDP capsule, the burn in the thicker HDC design takes place following the peak in average ion temperature.}
\label{HDCpdxp}
\end{center}
\end{figure}

\begin{figure}
\begin{center}
\includegraphics[width=0.9\textwidth]{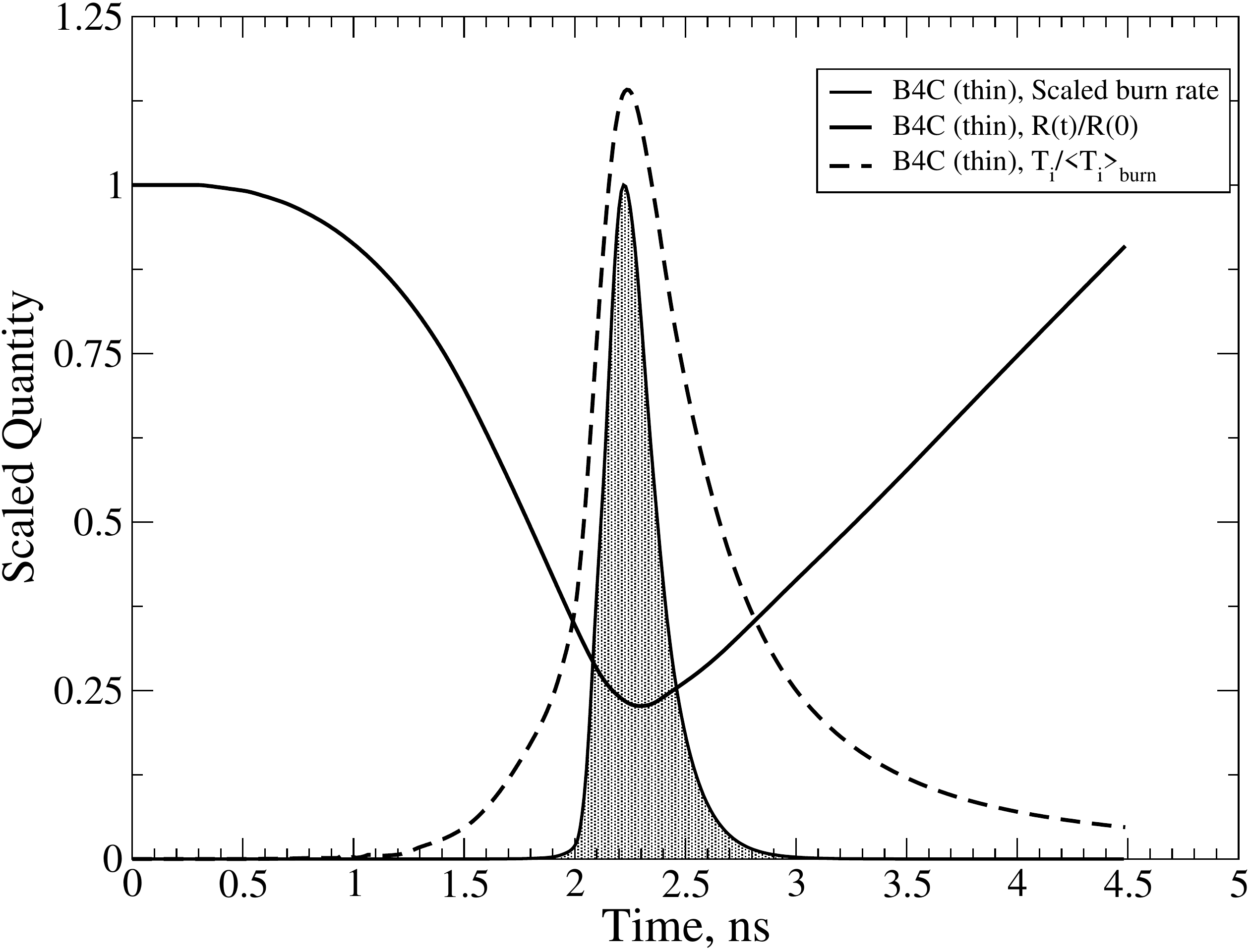}
\caption{Scaled burn rate (shaded curve), average radius (solid), and average ion temperature (dashed) as a function of time for the thinner B$_4$C design, as described in the text.  In contrast to the thicker HDC and GDP designs, the burn takes place primarily during the 
peak in the average gas ion temperature.  The compression in the thinner capsule design is also lower than it is for either the HDC or GDP thick capsules.}
\label{B4Cxp}
\end{center}
\end{figure}

In Figures~\ref{GDPpdxp}-\ref{B4Cxp} we plot several characteristic properties of the gas from simulations of the thicker GDP and HDC capsules 
(Figures~\ref{GDPpdxp} and \ref{HDCpdxp}), as well as the thinner B$_4$C capsule (Figure~\ref{B4Cxp}) as a function of time.  
In each of these plots, the burn rate is scaled by its peak value and the average ion temperature in the gas is scaled by the burn-averaged ion temperature listed in 
Table~\ref{yieldtable}.  We also plot the average radius of the gas scaled by the initial radius, which is equivalent to the convergence ratio as defined above and listed 
in Table~\ref{yieldtable}.  These plots demonstrate that the two thick capsule designs behave similarly, with most of the neutrons being produced after the peak in the 
average gas temperature, while the gas is still being compressed by the remaining ablator.  In contrast, the thin capsule design produces its yield at the same time as the ion 
temperature peaks.  The average ion temperature in the thin capsule design also exceeds the burn-averaged ion temperature, in contrast to the 
thick capsule designs.  In the thick designs, the burn is occurring primarily after shock convergence.  
This is 
consistent with what we found in our 1D study, as shown in Figures 2 and 7 of Ref.~\cite{Ellison18}, though in 2D the shock structure is more complicated and the burn 
is diminished mostly due to capsule break up, as opposed to capsule expansion, near peak compression.  The break up of the capsule occurs due to lower density 
regions that are generated at the points where the inner laser beams impact the capsule.

Comparing Figures~\ref{GDPpdxp} and \ref{HDCpdxp} provides some insight into why the HDC capsule produced a factor of 2 higher neutron yield than the GDP capsule.  
First, the increase in absorbed laser power for HDC relative to GDP leads to a stronger shock, and hence higher ion temperatures.  Second, the higher density ablator 
provides more remaining mass during stagnation, hence the capsule break up that leads to the demise of burn in the GDP calculation is less severe for HDC.  Third, the HDC 
implosion is slightly more symmetric than the GDP implosion.  The better symmetry and decreased breakup are evident in the scaled radius vs. time plot for HDC (Fig~\ref{HDCpdxp}), which shows a more obvious minimum at peak compression than the GDP capsule.

\begin{center}
\begin{figure}
\includegraphics[width=\textwidth]{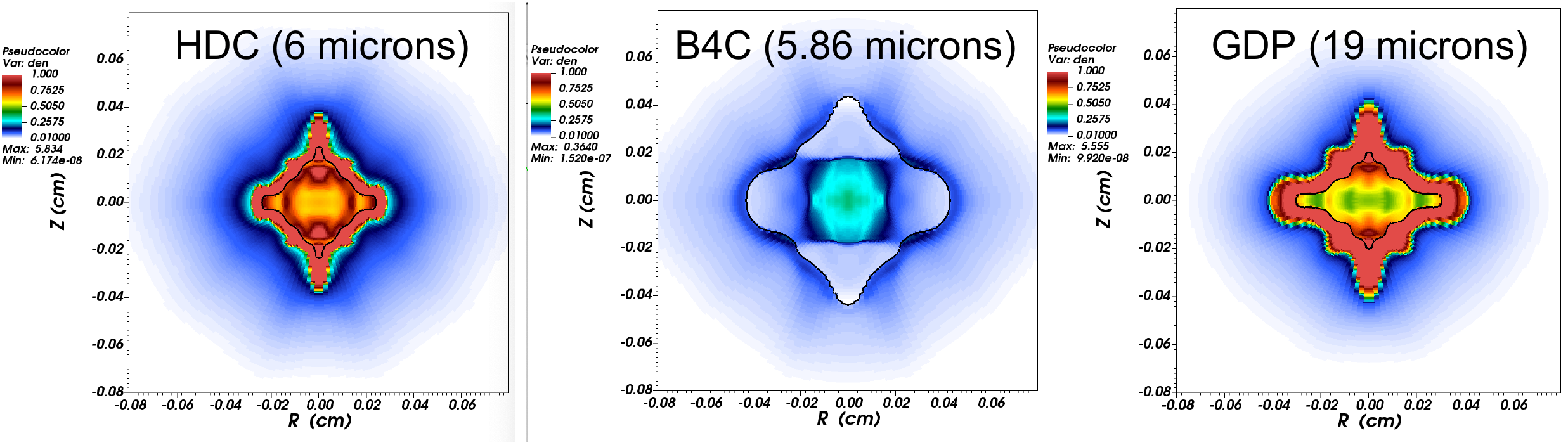}
\caption{Computed density profiles near peak compression for the HDC (left) and GDP (right) thicker capsule designs.  The thinner B$_4$C  (center) design 
shows significantly lower overall density than the other two capsules.  The color map is the same in all three images.  For the thinner B$_4$C design, 
the peak density of 0.36~g/cc occurs within the center of the gas and is surrounded by a region of lower density gas that extends out to about 400~$\mu$m 
in each direction.  The thicker ablators exhibit peak densities of $>5.5$~g/cc, and the peak density occurs in the remaining ablator that surrounds the gas.}
\label{implosions}
\end{figure}
\end{center}

Analogous to the similarity in the HDC and GDP thick capsules, we find that all of the thin capsule designs behave in a similar fashion, regardless of the identity of the ablator, 
producing similar yield, 
extremely high burn-averaged ion temperatures (15-16~keV) and about a factor of 2 lower convergence ratio than the thicker capsule designs.   
Figure~\ref{implosions} shows the density profile at peak compression for the B$_4$C thinner capsule design along with the computed density near peak 
compression for both the HDC and GDP thicker capsules.  The black contour in each plot is the boundary between the ablator and the gas.  
In the thinner capsules, the ablator has burned away, and the overall gas density is consequently 
lower than it is for the thicker capsules, consistent with the lower convergence shown in Fig.~\ref{B4Cxp}.  As discussed above, Figure~\ref{implosions} 
also shows that the HDC capsule produces a more uniform compression of the gas than the GDP ablator.  
All of the computed geometries at peak stagnation show significant asymmetry due to the 
polar direct drive configuration.  This indicates that the proposed heat flow experiments would still require significant design advances in order to realize a more 
uniform plasma, even with the use of a thinner ablator.  

\section{Summary and Conclusions}

In this paper we have presented a survey of candidate ablator materials for future experiments on the PDXP platform.    
Our simulations show that thinner capsule designs using 
the higher tensile strength materials should lead to more complete ablation of the capsule than the baseline GDP design.  
However, our calculations also show that, even in the case where a thinner ablator is used, the polar direct drive configuration is still predicted to imprint significant asymmetry on the implosion that persists through 
 stagnation.  Due to the inherent uncertainties in modeling direct drive implosions and the lack of experimental data to validate our models for ablators other than GDP, 
we have not yet pursued additional optimization of either the laser pointing or 
the laser pulse in this study.  It is possible that using a thinner ablator composed of any one of these materials and re-optimizing the drive for the new geometry would lead to a more symmetric implosion.  
Ideally, such optimization of the laser 
pulse would be performed using a baseline model that had been fit to experimental data for the actual ablator under consideration.  As such, we have saved this study as future work, and we hope that this work provides motivation for 
future experiments on this platform.

Even with the apparent lack of symmetry, the increased coupling of the laser to the higher density materials  improves the neutron yield in these implosions.  One interesting finding of this study is that 
an HDC version of N160920-005 is predicted to give about a factor of 2 higher neutron yield than the GDP ablator.  While the HDC capsule does give a slightly more symmetric geometry and somewhat higher gas density, 
the higher yield is primarily the result of better laser-target coupling, which produces a stronger shock and higher ion temperatures than the GDP capsule. 
Because the increase in yield can be largely attributed to better laser-capsule coupling, moving to an HDC capsule on the PDXP platform for neutron source 
development\cite{Yeamans18} appears to be a low risk change that would give higher yields than those that could be achieved in the current GDP-based experiments. 

While we have observed that the 2D model gives a reasonable fit to some of the diagnostic data obtained for N160920-005, modeling these 
direct drive simulations at the NIF is still relatively uncertain.  In the PDXP platform in particular, it is not clear what fraction of the yield is produced 
due to the strong shock versus what fraction comes from the compression of the gas by ablator material that remains after the laser pulse.  These calculations 
demonstrate large differences in the ablator areal density during the burn of the gas.  Ablators containing natural boron could enable us to determine whether 
the computed ablator areal density is realistic based on GRH measurements.  
Demonstrating the feasibility of those studies will require more detailed analysis of these simulations to determine whether there would be a 
measurable effect of remaining ablator mass on the GRH measurement.

\paragraph{Acknowledgments}  
This research was in part 
based upon work supported by the Department of Energy National Nuclear Security Administration under Award Number DE-NA0003856.
Part of the work was performed under the auspices of the U.S. Department of Energy by Lawrence Livermore National Laboratory under Contract No. DE-AC52-07NA27344.  H.~D. Whitley acknowledges the support of the PECASE award.  Contributions from the AWE are under the UK Ministry of Defence $\copyright$ Crown Owned Copyright 2020/AWE.

This document was prepared as an account of work sponsored by an agency of the United States government. Neither the United States government nor any agency thereof, nor any of their employees,
makes any warranty, express or implied, or assumes any legal
liability or responsibility for the accuracy, completeness, or
usefulness of any information, apparatus, product, or process
disclosed, or represents that its use would not infringe privately owned rights. Reference herein to any specific commercial product, process, or service by trade name, trademark,
manufacturer, or otherwise does not necessarily constitute
or imply its endorsement, recommendation, or favoring by
the U.S. Government or any agency thereof. The views and
opinions of authors expressed herein do not necessarily state
or reflect those of the U.S. Government or any agency thereof, and shall not be used for advertising or product endorsement purposes.

\bibliography{pdxp.bib}

\begin{thebibliography}{10}
\expandafter\ifx\csname url\endcsname\relax
  \def\url#1{\texttt{#1}}\fi
\expandafter\ifx\csname urlprefix\endcsname\relax\def\urlprefix{URL }\fi
\expandafter\ifx\csname href\endcsname\relax
  \def\href#1#2{#2} \def\path#1{#1}\fi

\bibitem{Ellison18}
C.~L. Ellison, H.~D. Whitley, C.~R.~D. Brown, S.~R. Copeland, W.~J. Garbett,
  H.~P. Le, M.~B. Schneider, Z.~B. Walters, H.~Chen, J.~I. Castor, R.~S.
  Craxton, M.~Gatu~Johnson, E.~M. Garcia, F.~R. Graziani, G.~E. Kemp, C.~M.
  Krauland, P.~W. McKenty, B.~Lahmann, J.~E. Pino, M.~S. Rubery, H.~A. Scott,
  R.~Shepherd, H.~Sio, Development and modeling of a polar direct-drive
  exploding pusher platform at the {N}ational {I}gnition {F}acility, Physics of
  Plasmas 25~(7) (2018) 072710.
\newblock \href {http://dx.doi.org/10.1063/1.5025724}
  {\path{doi:10.1063/1.5025724}}.

\bibitem{Miles_2012}
A.~R. Miles, H.-K. Chung, R.~Heeter, W.~Hsing, J.~A. Koch, H.-S. Park, H.~F.
  Robey, H.~A. Scott, R.~Tommasini, J.~Frenje, C.~K. Li, R.~Petrasso,
  V.~Glebov, R.~W. Lee, Numerical simulation of thin-shell direct drive
  {DH}e3-filled capsules fielded at {OMEGA}, Physics of Plasmas 19~(7) (2012)
  072702.
\newblock \href {http://dx.doi.org/10.1063/1.4737052}
  {\path{doi:10.1063/1.4737052}}.

\bibitem{Skupsky04}
S.~Skupsky, J.~A. Marozas, R.~S. Craxton, R.~Betti, T.~J.~B. Collins, J.~A.
  Delettrez, V.~N. Goncharov, P.~W. McKenty, P.~B. Radha, T.~R. Boehly, J.~P.
  Knauer, F.~J. Marshall, D.~R. Harding, J.~D. Kilkenny, D.~D. Meyerhofer,
  T.~C. Sangster, R.~L. McCrory, Polar direct drive on the national ignition
  facility, Physics of Plasmas 11~(5) (2004) 2763--2770.
\newblock \href {http://dx.doi.org/10.1063/1.1689665}
  {\path{doi:10.1063/1.1689665}}.

\bibitem{GatuJohnson18}
M.~Gatu~Johnson, D.~T. Casey, M.~Hohenberger, A.~B. Zylstra, A.~Bacher, C.~R.
  Brune, R.~M. Bionta, R.~S. Craxton, C.~L. Ellison, M.~Farrell, J.~A. Frenje,
  W.~Garbett, E.~M. Garcia, G.~P. Grim, E.~Hartouni, R.~Hatarik, H.~W.
  Herrmann, M.~Hohensee, D.~M. Holunga, M.~Hoppe, M.~Jackson, N.~Kabadi, S.~F.
  Khan, J.~D. Kilkenny, T.~R. Kohut, B.~Lahmann, H.~P. Le, C.~K. Li, L.~Masse,
  P.~W. McKenty, D.~P. McNabb, A.~Nikroo, T.~G. Parham, C.~E. Parker, R.~D.
  Petrasso, J.~Pino, B.~Remington, N.~G. Rice, H.~G. Rinderknecht, M.~J.
  Rosenberg, J.~Sanchez, D.~B. Sayre, M.~E. Schoff, C.~M. Shuldberg, F.~H.
  Séguin, H.~Sio, Z.~B. Walters, H.~D. Whitley, Optimization of a high-yield,
  low-areal-density fusion product source at the {N}ational {I}gnition
  {F}acility with applications in nucleosynthesis experiments, Physics of
  Plasmas 25~(5) (2018) 056303.
\newblock \href {http://dx.doi.org/10.1063/1.5017746}
  {\path{doi:10.1063/1.5017746}}.

\bibitem{Yeamans18}
C. B. Yeamans and B. E. Blue, National Ignition Facility neutron sources,
  Technical Report No. LLNL-CONF-739397, Lawrence Livermore National
  Laboratory, 2018.
\newblock \href{https://www.osti.gov/servlets/purl/1458648}{[link]}.
\newline\urlprefix\url{https://www.osti.gov/servlets/purl/1458648}

\bibitem{Yeamans20}
C.~Yeamans, G.~E. Kemp, Z.~B. Walters, P.~W. McKenty, E.~M. Garcia, Y.~Yang,
  R.~S. Craxton, B.~E. Blue, High yield polar direct drive fusion neutron
  sources at the {N}ational {I}gnition {F}acility, {\em Submitted to Nuclear
  Fusion}.

\bibitem{Nikroo04}
A.~Nikroo, D.~G. Czechowicz, K.~C. Chen, M.~Dicken, C.~Morris, R.~Andrews,
  A.~Greenwood, E.~Castillo, Mechanical properties of thin {GDP} shells used as
  cryogenic direct drive targets at {OMEGA}, Fusion Science and Technology
  45~(2) (2004) 229--232.
\newblock \href {http://dx.doi.org/10.13182/FST45-2-229}
  {\path{doi:10.13182/FST45-2-229}}.

\bibitem{specialWhitley}
The shots in September 2016 were designed to provide calibration data for the
  2D simulations and examine whether symmetry could be improved via
  modifications in the cone fraction. As such, we tested three different cone
  fractions (0.33, 0.28, and 0.22, respectively for N160920-003,N160920-005,
  and N160921-003.) We chose to use N160920-005 for this study because it
  showed the best in-flight symmetry. More complete results from these shots
  will be included in a future report.

\bibitem{Rosen79}
M.~D. Rosen, J.~H. Nuckolls, Exploding pusher performance: A theoretical model,
  The Physics of Fluids 22~(7) (1979) 1393--1396.
\newblock \href {http://dx.doi.org/10.1063/1.862752}
  {\path{doi:10.1063/1.862752}}.

\bibitem{GRHref}
H.~Geppert-Kleinrath, Y.~Kim, K.~D. Meaney, H.~W. Herrmann, N.~M. Hoffman,
  J.~A. Carrera, A.~L. Kritcher, M.~S. Rubery, A.~Leatherland, High bandwidth
  {DT} reaction history measurements in inertial confinement fusion, Bulletin
  of the American Physical Society Vol. 64.

\bibitem{GRH1}
D.~B. Sayre, L.~A. Bernstein, J.~A. Church, H.~W. Herrmann, W.~Stoeffl,
  Multi-shot analysis of the gamma reaction history diagnostic, Review of
  Scientific Instruments 83~(10) (2012) 10D905.
\newblock \href {http://dx.doi.org/10.1063/1.4729492}
  {\path{doi:10.1063/1.4729492}}.

\bibitem{Hayes15}
A.~C. Hayes-Sterbenz, G.~M. Hale, G. Jungman, and M.~W. Paris, Probing the
  Physics of Burning {DT} Capsules Using Gamma-ray Diagnostics, Technical
  Report No. LA-UR-15-20627, Los Alamos National Laboratory, 2015.
\newblock \href {http://dx.doi.org/10.2172/1169150}
  {\path{doi:10.2172/1169150}}.

\bibitem{Hayes17}
A.~C. Hayes, Applications of nuclear physics, Reports on Progress in Physics
  80~(2) (2017) 026301.
\newblock \href {http://dx.doi.org/10.1088/1361-6633/80/2/026301}
  {\path{doi:10.1088/1361-6633/80/2/026301}}.

\bibitem{specialHayes}
Anna Sterbenz-Hayes, private communication.

\bibitem{GRH2}
K.~D. Meaney, Y.~H. Kim, H.~W. Herrmann, H.~Geppert-Kleinrath, N.~M. Hoffman,
  Improved inertial confinement fusion gamma reaction history 12c gamma-ray
  signal by direct subtraction, Review of Scientific Instruments 90~(11) (2019)
  113503.
\newblock \href {http://dx.doi.org/10.1063/1.5092501}
  {\path{doi:10.1063/1.5092501}}.

\bibitem{Chen17}
R.~Chen, Q.~Shi, L.~Su, M.~Yang, Z.~Huang, Y.~Shi, Q.~Zhang, Z.~Liao, T.~Lu,
  Preparation of a b4c hollow microsphere through gel-casting for an inertial
  confinement fusion ({ICF}) target, Ceramics International 43~(1, Part A)
  (2017) 571 -- 577.
\newblock \href {http://dx.doi.org/10.1016/j.ceramint.2016.09.196}
  {\path{doi:10.1016/j.ceramint.2016.09.196}}.

\bibitem{Dufrane16}
W.~D. Frane, O.~Cervantes, G.~Ellsworth, J.~Kuntz, Consolidation of cubic and
  hexagonal boron nitride composites, Diamond and Related Materials 62 (2016)
  30 -- 41.
\newblock \href {http://dx.doi.org/10.1016/j.diamond.2015.12.003}
  {\path{doi:10.1016/j.diamond.2015.12.003}}.

\bibitem{Zhang2018b}
S.~Zhang, B.~Militzer, M.~C. Gregor, K.~Caspersen, L.~H. Yang, J.~Gaffney,
  T.~Ogitsu, D.~Swift, A.~Lazicki, D.~Erskine, R.~A. London, P.~M. Celliers,
  J.~Nilsen, P.~A. Sterne, H.~D. Whitley, Theoretical and experimental
  investigation of the equation of state of boron plasmas, Phys. Rev. E 98
  (2018) 023205.
\newblock \href {http://dx.doi.org/10.1103/PhysRevE.98.023205}
  {\path{doi:10.1103/PhysRevE.98.023205}}.

\bibitem{Zhang2019BN}
S.~Zhang, A.~Lazicki, B.~Militzer, L.~H. Yang, K.~Caspersen, J.~A. Gaffney,
  M.~W. D{\"{a}}ne, J.~E. Pask, W.~R. Johnson, A.~Sharma, P.~Suryanarayana,
  D.~D. Johnson, A.~V. Smirnov, P.~A. Sterne, D.~Erskine, R.~A. London,
  F.~Coppari, D.~Swift, J.~Nilsen, A.~J. Nelson, H.~D. Whitley, {Equation of
  state of boron nitride combining computation, modeling, and experiment},
  Physical Review B 99~(16) (2019) 165103.
\newblock \href {http://dx.doi.org/10.1103/PhysRevB.99.165103}
  {\path{doi:10.1103/PhysRevB.99.165103}}.

\bibitem{Zhang20}
S.~Zhang, M.~C. Marshall, L.~H. Yang, P.~A. Sterne, B.~Militzer, M.~D\"ane,
  J.~A. Gaffney, A.~Shamp, T.~Ogitsu, K.~Caspersen, A.~E. Lazicki, D.~Erskine,
  R.~A. London, P.~M. Celliers, J.~Nilsen, H.~D. Whitley, Benchmarking boron
  carbide equation of state using computation and experiment, Phys. Rev. E 102
  (2020) 053203.
\newblock \href {http://dx.doi.org/10.1103/PhysRevE.102.053203}
  {\path{doi:10.1103/PhysRevE.102.053203}}.

\bibitem{Sterne_2016}
P.~A. Sterne, L.~X. Benedict, S.~Hamel, A.~A. Correa, J.~L. Milovich, M.~M.
  Marinak, P.~M. Celliers, D.~E. Fratanduono, Equations of state for ablator
  materials in inertial confinement fusion simulations, Journal of Physics:
  Conference Series 717 (2016) 012082.
\newblock \href {http://dx.doi.org/10.1088/1742-6596/717/1/012082}
  {\path{doi:10.1088/1742-6596/717/1/012082}}.

\bibitem{Zhang2018}
S.~Zhang, B.~Militzer, L.~X. Benedict, F.~Soubiran, P.~A. Sterne, K.~P. Driver,
  Path integral {M}onte {C}arlo simulations of dense carbon-hydrogen plasmas,
  J. Chem. Phys. 148~(10) (2018) 102318.
\newblock \href {http://dx.doi.org/10.1063/1.5001208}
  {\path{doi:10.1063/1.5001208}}.

\bibitem{Benedict2014}
L.~X. Benedict, K.~P. Driver, S.~Hamel, B.~Militzer, T.~Qi, A.~A. Correa,
  A.~Saul, E.~Schwegler, Multiphase equation of state for carbon addressing
  high pressures and temperatures, Phys. Rev. B 89 (2014) 224109.
\newblock \href {http://dx.doi.org/10.1103/PhysRevB.89.224109}
  {\path{doi:10.1103/PhysRevB.89.224109}}.

\bibitem{Darlington01}
R.~M. Darlington, T.~L. McAbee, G.~Rodrigue, A study of {ALE} simulations of
  rayleigh-taylor instability, Computer Physics Communications 135~(1) (2001)
  58 -- 73.
\newblock \href {http://dx.doi.org/10.1016/S0010-4655(00)00216-2}
  {\path{doi:10.1016/S0010-4655(00)00216-2}}.

\bibitem{Morgan16}
B.~E. Morgan, J.~A. Greenough, Large-eddy and unsteady {RANS} simulations of a
  shock-accelerated heavy gas cylinder, Shock Waves 26~(4) (2016) 355--383.
\newblock \href {http://dx.doi.org/10.1007/s00193-015-0566-3}
  {\path{doi:10.1007/s00193-015-0566-3}}.

\bibitem{Krasheninnikova_2014}
N.~S. Krasheninnikova, J.~A. Cobble, T.~J. Murphy, I.~L. Tregillis, P.~A.
  Bradley, P.~Hakel, S.~C. Hsu, G.~A. Kyrala, K.~A. Obrey, M.~J. Schmitt,
  et~al., Designing symmetric polar direct drive implosions on the {O}mega
  {L}aser {F}acility, Physics of Plasmas 21~(4) (2014) 042703.
\newblock \href {http://dx.doi.org/10.1063/1.4870756}
  {\path{doi:10.1063/1.4870756}}.

\bibitem{Murphy_2015}
T.~J. Murphy, N.~S. Krasheninnikova, G.~Kyrala, P.~A. Bradley, J.~A.
  Baumgaertel, J.~Cobble, P.~Hakel, S.~C. Hsu, J.~L. Kline, D.~Montgomery,
  et~al., Laser irradiance scaling in polar direct drive implosions on the
  {N}ational {I}gnition {F}acility, Physics of Plasmas 22~(9) (2015) 092707.
\newblock \href {http://dx.doi.org/10.1063/1.4931092}
  {\path{doi:10.1063/1.4931092}}.

\bibitem{Radha16}
P.~Radha, M.~Hohenberger, D.~Edgell, J.~Marozas, F.~Marshall, D.~Michel,
  M.~Rosenberg, W.~Seka, A.~Shvydky, T.~Boehly, et~al., Direct drive:
  Simulations and results from the {N}ational {I}gnition {F}acility, Physics of
  Plasmas 23~(5) (2016) 056305.
\newblock \href {http://dx.doi.org/10.1063/1.4946023}
  {\path{doi:10.1063/1.4946023}}.

\bibitem{Zylstra20}
A.~B. Zylstra, C.~Yeamans, S.~Le~Pape, A.~MacKinnon, M.~Hohenberger, D.~N.
  Fittinghoff, H.~Herrmann, Y.~Kim, P.~B. Radha, P.~W. McKenty, R.~S. Craxton,
  M.~Hoppe, Enhanced direct-drive implosion performance on nif with wavelength
  separation, Physics of Plasmas 27~(12) (2020) 124501.
\newblock \href {http://dx.doi.org/10.1063/5.0021015}
  {\path{doi:10.1063/5.0021015}}.

\end{thebibliography}

\end{document}